\documentclass[twocolumn,aps,showpacs,showkeys,floatfix]{revtex4}
\usepackage{amssymb,amsmath}
\usepackage[pdftex]{graphicx}

\renewcommand{\epsilon }{\varepsilon}
\renewcommand{\phi}{\varphi}

\def\be{\begin{equation}}
\def\ee{\end{equation}}
\def\bq{\begin{eqnarray}}
\def\eq{\end{eqnarray}}
\def\beq{\begin{eqnarray}}
\def\eeq{\end{eqnarray}}

\begin{document} 


\title{Representing perturbed dynamics in biological network models}
\author{Gautier Stoll$^{1,3}$}
\email{Gautier.stoll@curie.fr}
\author{Jacques Rougemont$^{3}$}
\email{jacques.rougemont@isb-sib.ch}
\author{Felix Naef$^{1,2,3}$}
\email{felix.naef@isrec.ch}

\affiliation{$^1$NCCR Molecular Oncology, ch. des Boveresses 155, 1066
  Epalinges, Switzerland} 
\affiliation{$^3$School of Life Sciences, ISREC, Ecole polytechnique
  F\'ed\'erale de Lausanne 1015 Lausanne, Switzerland} 
\affiliation{$^3$Swiss Institute of Bioinformatics, Quartier Sorge-Genopode, 1015
  Lausanne, Switzerland} 

\begin{abstract}
We study the dynamics of gene activities in relatively small size biological networks (up to a few tens of nodes), e.g. the 
activities of cell-cycle proteins during the mitotic cell-cycle progression.
Using the framework of deterministic discrete dynamical models, we characterize the dynamical modifications in response
to structural perturbations in the network connectivities.
In particular, we focus on how perturbations affect the set of fixed points and sizes of the basins of attraction.
Our approach uses two analytical measures: the basin entropy $H$ and the perturbation size $\Delta$, a quantity that reflects
the distance between the set of fixed points of the perturbed network to that of
the unperturbed network.
Applying our approach to the yeast-cell cycle network introduced by
Li \textit{et al.}
provides a low dimensional and informative fingerprint of network behavior under
large classes of perturbations. We identify
interactions that are crucial for proper network function, and also pinpoints
functionally redundant network connections. 
Selected perturbations exemplify the breadth of dynamical responses in this cell-cycle model.
\end{abstract}

\maketitle


\section{Introduction}  

Recent experimental developments in the fields of genomics, e.g. whole genome DNA sequencing or proteomics,
are opening possibilities for systems level studies in biology \cite{leibler_3,arkin_4,barabasi_3,alon_8}.
In particular, the notion that biological functions may rely on a large number of interconnected variables (for example genes) working in concert has stimulated general theoretical interest
about properties of biological networks \cite{barabasi_2}.
Studies of the statistical properties of large (typically thousands of nodes) biological networks have identified
a number of functional building block, termed network motifs, that occur more frequently than random \cite{alon_9}.
These findings support the idea that some systems are designed around a modular architecture, in which autonomous modules are wired
together to generate versatile biological functions \cite{leibler_3,barkai_9,alon_8,eckmann}.
While structural (or topological) properties are key for network characterization, functional properties are ultimately
encoded in dynamical, or time-dependent changes in the state variables of the nodes.
The sizes of systems that can be modeled dynamically
are typically much smaller (10-100 nodes). One common modeling approach, for example for the yeast cell-cycle \cite{tyson_novak_5},
is to simulate the nonlinear system of chemical rate equations describing the putative biochemical processes.
Modeling approaches have been applied to a number of systems,
including the cell-cycle \cite{tyson_novak_5,cross_3}, the lambda-phage switch in E. coli \cite{arkin2_4}.
Although these models provide a detailed description, this approach suffers from the caveat that most parameters are currently not accessible experimentally.
In addition, the number of parameters is typically about five per reaction,
resulting in a prohibitively large parameter space.
This last point makes it difficult to grasp the full solution space of the model. Recent approaches based on
sampling the parameter space in optimal regions have been developed \cite{brown_sethna_1}.
At the opposite end of model complexity, dynamical rules based on boolean state variables have been useful
for studying more global dynamical properties of topological classes of networks \cite{kadanoff,kauffman_1}.
In addition, boolean models have been successfully applied to the yeast cell-cycle \cite{tang_1, stoll} and the body patterning
in drosophila embryos \cite{albert_3,albert_4}.

In this study, we develop a systematic approach to describe how the dynamical
landscape of small (less than about 50 nodes) boolean networks is affected by perturbations in the network connectivity.
In particular, we consider the basin entropy $H$, a quantity that considers the size distribution of
the basins of attraction.
We complement entropy with a measure of distance between the stable
fixed points of a perturbed network and those in the unperturbed network. This combination
gives a low-dimensional and compact representation of the patterns induced by a
large number of perturbations. We illustrate our methods using the yeast cell-cycle network
introduced in \cite{tang_1}, and discuss examples of structural perturbations producing a range
of modified basins of attraction. 


\section{Definitions}  
Following \cite{tang_1} a network of $N$ nodes can be represented by a $N\times N$ adjacency matrix $A$, in which an activating link between node $i$ and node $j$ is 
represented by $A_{ij}=1$ and an inhibiting link by $A_{ij}=-1$.
The possibility of self-inhibitory (or activating links) $A_{ii}=\pm1$ is not excluded.
In the Boolean approximation, each node has two possible states,
so that the global state of all nodes can be represented by a vector
$\mathbf{S}$, with $S_i=1$ when the node $i$ is \textit{on} and $S_i=0$ if
the node is \textit{off}.
The full phase space containing $2^N$ states is denoted by $\Lambda$.

\subsection{Boolean dynamics} A simple dynamical rule that
characterizes the temporal evolution of the state variable can be
defined following \cite{tang_1}, which is closely related to update
rules applied in perceptron models. 
If the network is in the state $\mathbf{S}(t)$ at time $t$, the 
state at the next time-step $\mathbf{S}(t+1)$ is given by:

\bq
S_i(t+1)=\left\{\begin{array}{ccc}
1 &\text{if}&\sum_j A_{ij}S_j(t)>0\\
S_i(t)&\text{if}&\sum_j A_{ij}S_j(t)=0\\
0 &\text{if}&\sum_j A_{ij}S_j(t)<0\\
\end{array}\right.
\eq

For a given network, we apply this rule to every possible initial
condition in $\Lambda$. This defines orbits (trajectories) that must
end in a limit cycle (periodic attractor) since we are dealing with a
dynamical system on a finite space. A fixed point is a cycle of length one.

Accordingly, $\Lambda$ can be decomposed into a disjoint union of $K$ basins
of attraction $B_k$ of size $d_k$: $\Lambda=\bigcup_{k=1}^KB_k$.

In a biological network, the attractors correspond to functional endpoints, and it is important that the states in the attractors are 
consistent with observed data.
For example, by far the largest endpoint in the cell-cycle network of
Li et al.~(see appendix) corresponds to the stationary G1 phase in the cycle.
Other systems are more switch-like, for instance in signal transduction,
where a cell might change its state from growth to differentiation according 
to an external trigger.
To characterize these attractors, we introduce the following definitions:

\begin{itemize}
\item
We compute the \textit{number of attractors} $K$: an
attractor is a limit cycle or a fixed point. An attractor $A$ has a basin of
attraction $B$ which is the set of all initial conditions whose orbit
converges to $A$.

\hspace{-1cm}
\item
The \textit{basin entropy} $H$ is defined as follows:
let $p_k=2^{-N}d_k$ be the probability that an initial state belongs
to basin $B_k$. Then, the entropy reads 
\be
H:=-\sum_{k=1}^K p_k\,\log\left(p_k\right)
\ee
$H$ is maximum ($H=\log(K)$) if each state is its own basin of size
one, and minimum ($H=0$) when there is one single basin.
$H$ is a natural measure for characterizing basin structures \cite{book}.
Because it takes into account the relative basin sizes, it is quite
insensitive to appearance of small and biologically irrelevant
basins. 

\item The \textit{perturbation size} $\Delta$
measures the distance between attractors of a perturbed and a reference
network: from every initial conditions, 
the Hamming distance between the fixed points is computed, and the
average over all initial conditions is taken. More precisely, if
$\mathbf{FP}_G(\mathbf{S})$ 
is the fixed point of the trajectory starting at $\mathbf{S}$ and
generated by the network G, then 
\begin{equation}
\Delta_{G,G'}:=\frac{1}{2^N}\sum_{\mathbf{S}}
\mbox{HAM}(\mathbf{FP}_G(\mathbf{S}),\mathbf{FP}_{G'}(\mathbf{S}))
\label{Delta_def}
\end{equation}
where $\mbox{HAM}(\cdot,\cdot)$ 
is the Hamming distance between two boolean states, namely
\begin{equation}
\mbox{HAM}(\mathbf{S},\mathbf{T}):=\frac{1}{N}\sum_i |S_i-T_i|
\end{equation}

The value $\Delta$ has the following interpretation: it is the average
probability (taken over all nodes)  that, for a random initial
condition, the final state of a node differs.
In this study, the reference network $G$ will be the cell-cycle network of
Li et al., which has one very a large basin of attraction and several
smaller ones. If some trajectories in the perturbed networks $G'$ end in 
a limit cycle, $\Delta$ is defined as the average of the Hamming distance
along the cycle.


\end{itemize}


\subsection{Network models and perturbations}

Our goal is to assess how network dynamics is affected by several types of
perturbations.  We consider two classes: one which randomizes the adjacency matrix
while keeping a number of topological characteristics from the original network invariant.
The second class mimics biological perturbations, as would occur for example through
mutations in the interaction partners that constitute the network links.The two classes are defined as follows:

\begin{itemize}
\item \textit{Shuffle} (class I): all activating and inhibiting arrows are cut in half and re-wired randomly. This ensure that the connectivity at each node is conserved.
As compared to the Li et al.~\cite{tang_1} study, we generate random networks that are more constrained, since the
connectivity at each node is forced to remain unchanged after randomization.
Such perturbations are applied in the studies of network motifs
\cite{alon_8,alon_9}.

\item \textit{Remove} (class II): the arrows are simply suppressed. We extend this class of perturbations beyond single link removal.

\end{itemize}


\section{Results}

We study the yeast cell-cycle network of Li et al.~\cite{tang_1}
(the \textit{Yeast cell-cycle network} or \textit{YCC}),
in which a boolean model reproducing the different phases of
the cycle is constructed (see appendix).
This model has a main fixed point attracting $86\%$ of the intial
conditions.
Biologically this state corresponds to the G1 stationary phase of the
cell-cycle, as reflected by the activities of the respective nodes.
Using computer simulations,
the authors further showed that the cell-cycle dynamics had certain
robustness properties when challenged with perturbations.
In particular, it was shown that in a majority of cases,
removal of one link or addition of a link at random did not change much the size of the largest basin of attraction.
Finally, the studied network had unusual trajectory channeling
properties, when compared to random networks with equal number
of nodes and links.
Here we extend the characterization of this model by introducing a
combination of measures to characterize the structure of basins of
attraction as they are modified by structural perturbations. In particular we
investigate the consequences of combined mutations and show that they can lead
to cancellation effect.

\subsection{Study of shuffled networks (Class I perturbations)}

This type of perturbation allows to study the dynamical characteristics
of a biological network in comparison with random networks belonging to a
topological class.
Figure 1 shows the Number of attractors ($K$) and the Entropy ($H$) of the YCC
and randomly shuffled (Class I) versions thereof. 

\begin{figure}[h]
\includegraphics[width=9cm,angle=0]{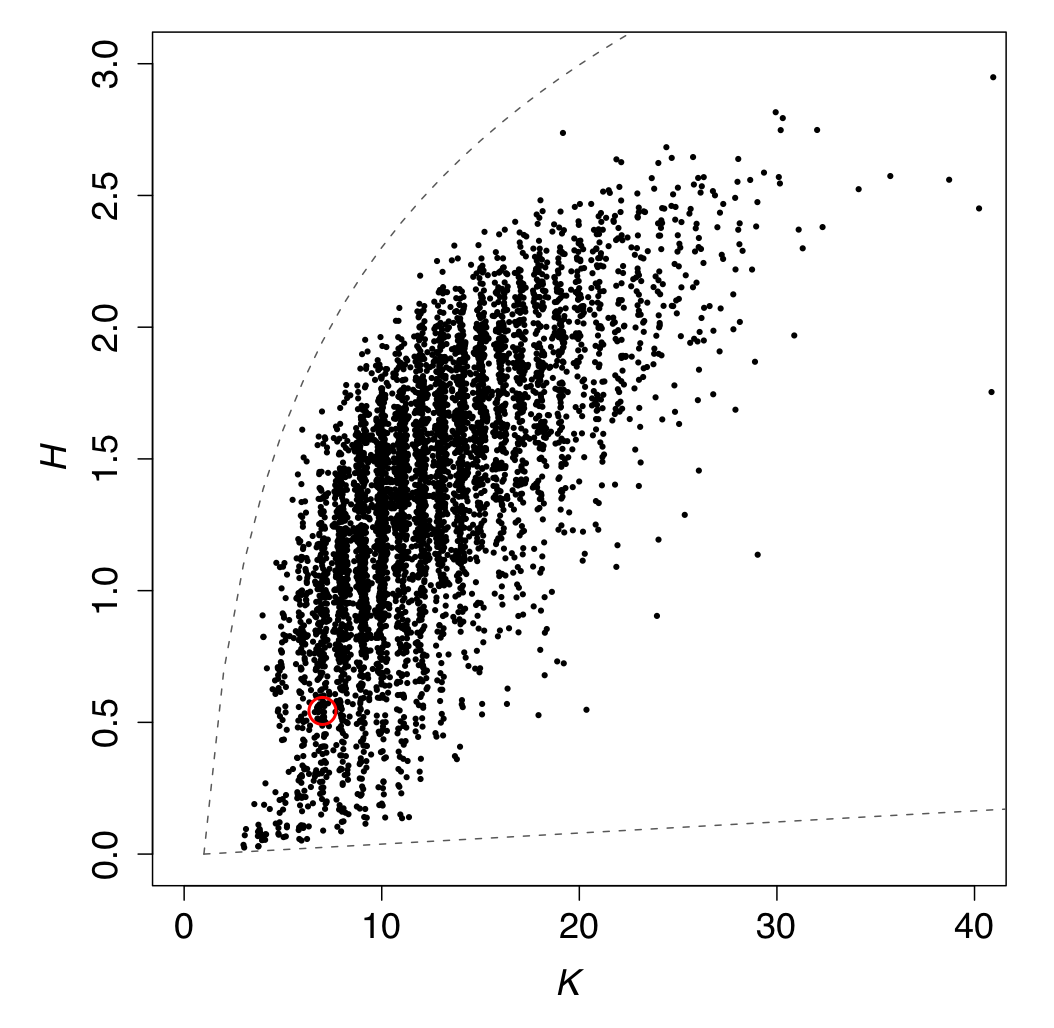}
\label{fig_shuff}
\caption{Entropy vs. number of attractors after class I perturbation (shuffled arrows). The range of possible $H$ values is indicated by the dashed gray lines. The open red circle represents the reference network, the other points 
show the perturbed networks.}
\end{figure}

The location of the reference network in the $H-K$
plane respective to the scatter of the perturbed networks allows us to asses how
typical a network behaves with respect to a class.
Accordingly, the YCC is atypical, as seen by its marginal location in the lower left
corner. Indeed, this network has lower entropy and fewer basins than most networks, consistent
with \cite{tang_1}.

\subsection{Study of mutated networks (Class II perturbations)}

The previous discussion shows how entropy characterizes the system
of attractors. However, $H$ contains only information about the relative
weights of the attractors, irrespective of their biological relevance.
For example a perturbation can decrease the entropy while shifting the
fixed point away from from that in the unperturbed, biologically
relevant state.
For this reason we introduced a second quantity, $\Delta$ (Equation
\ref{Delta_def}), a probabilistic measure of the change in the
fixed point after perturbation. Therefore, $\Delta$ reflects the
change in the biological relevance of the basin structure.

\begin{figure}[h]
\centering
\includegraphics[width=4cm,angle=0]{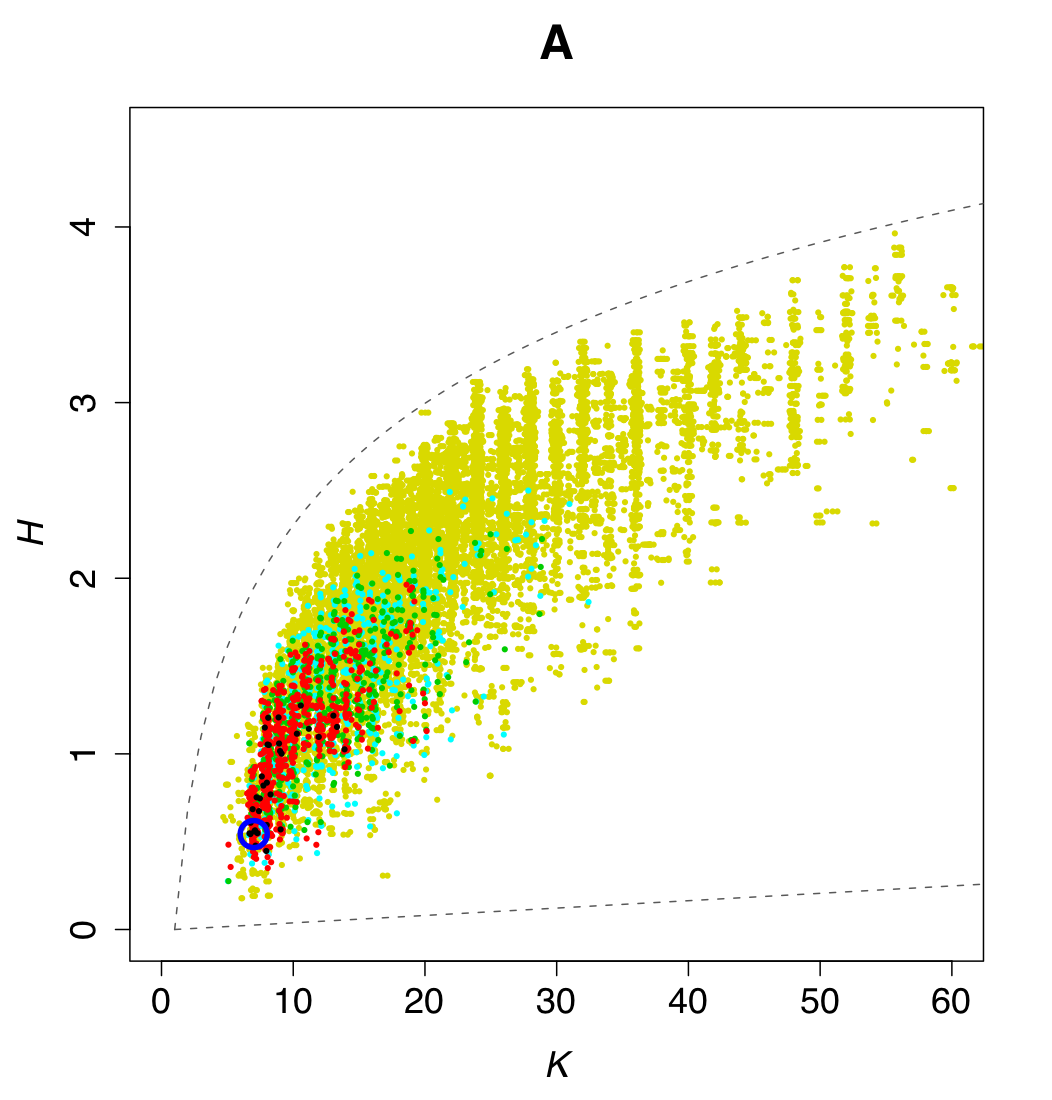}
\includegraphics[width=3.8cm,angle=0]{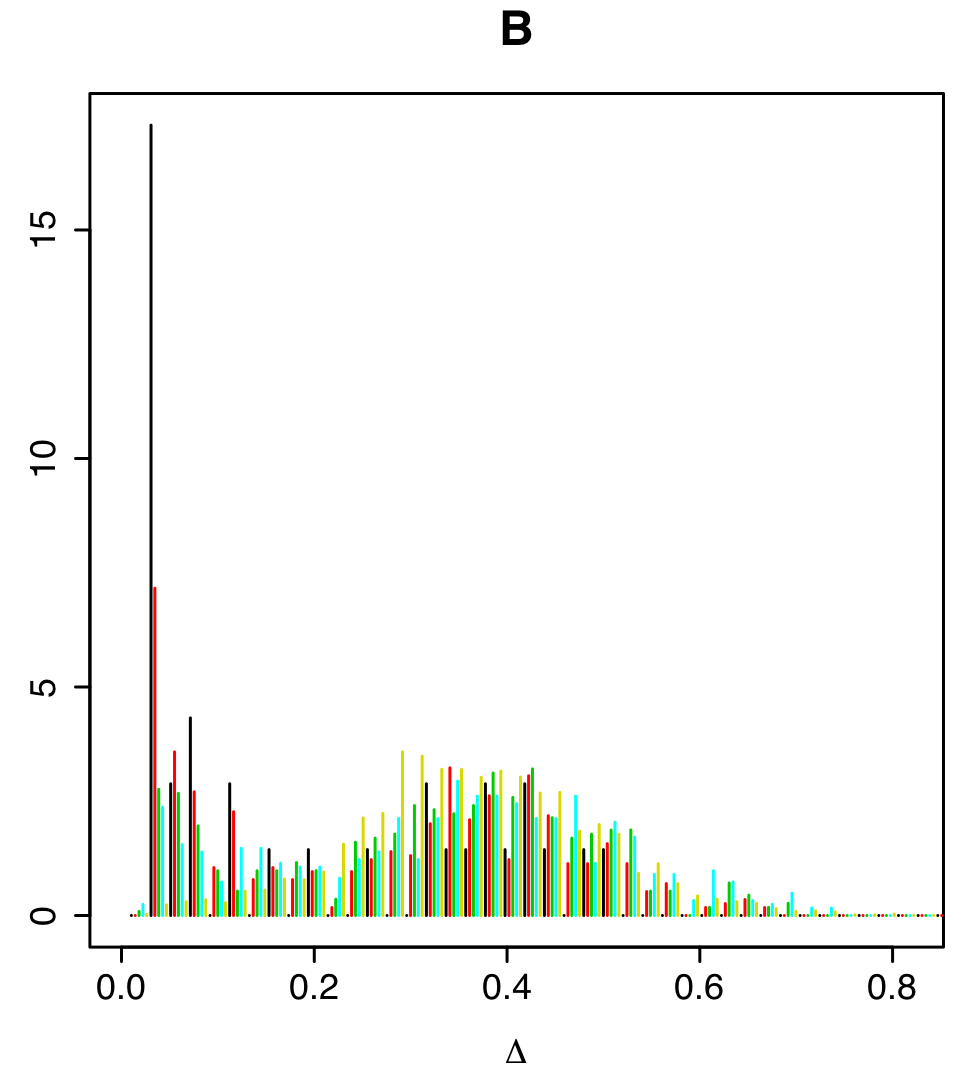}
\includegraphics[width=9cm,angle=0]{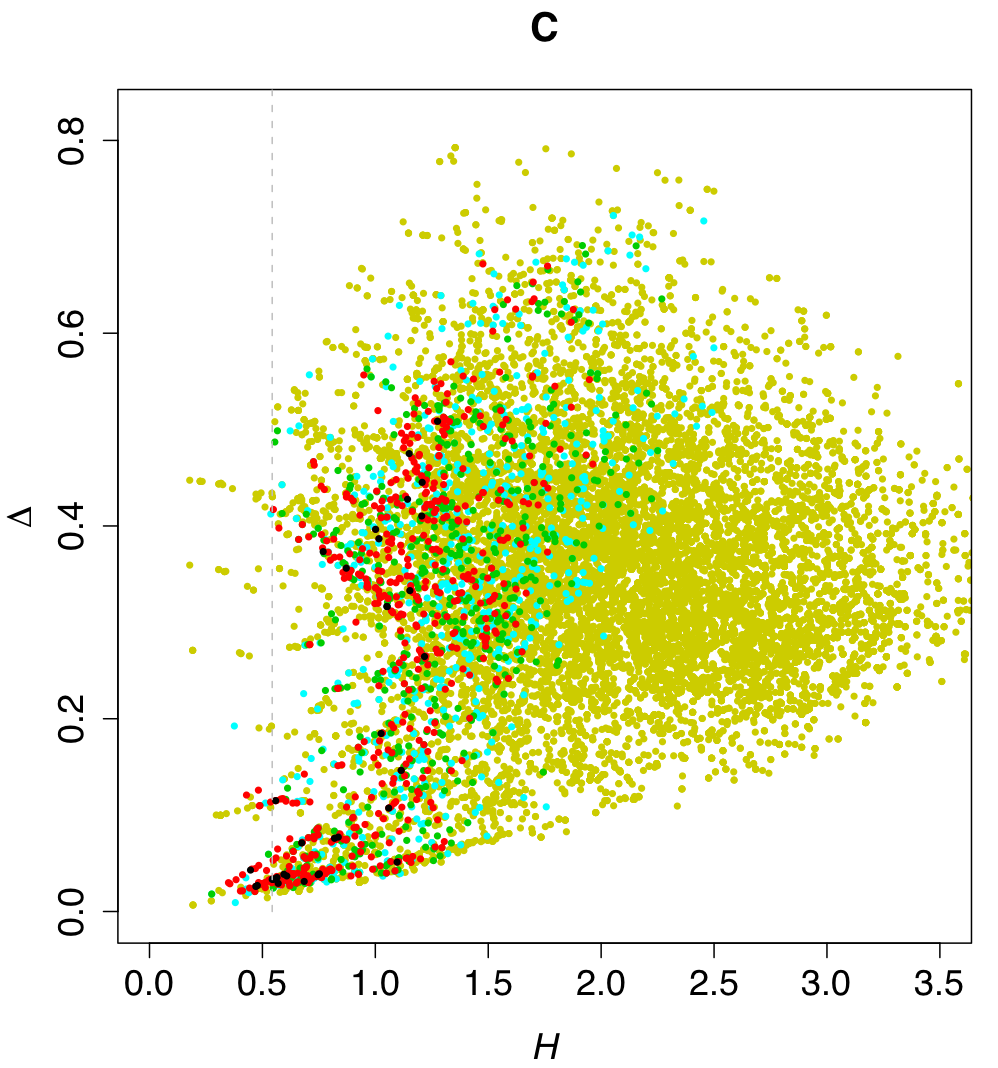}
\label{fig_rmv}
\caption{Entropy, number of attractors and $\Delta$ after class II perturbation
(removed arrows). Colors represent different number of removed arrows:
black for one removed arrows, red 
for 2, green for 3, turquoise for 4 and yellow for more than 4.
A: same figure as for the class I perturbation,
the range of possible $H$ values is indicated by the dashed gray lines,
the open blue circles represent the reference network.
B: Distribution of $\Delta$.
C: $\Delta$ vs. $H$ plot, the dashed gray line represents the
entropy of the reference network.}
\end{figure}

We first repeat Figure 1 for class II perturbations which shows that networks with few perturbations
cluster around the wild-type model (Figure 2A), while the sread for networks with four perturbations resembles
the shuffled models (Figure 1). Turning to the measure of $\Delta$, we find that
$\Delta$-distribution
(Figure 2B) is bimodal, showing two distinct populations of perturbations:
($\Delta\lesssim 0.2$ and $\Delta\gtrsim 0.2$).
In the second case, the perturbed
model does not reproduce the biologically correct cell-cycle progression.
But if $\Delta$ is small, then the system of attractors of the perturbed network
is still consistent with the biology and entropy allows to
discriminate between networks with a larger or smaller main basin of attraction.
For this reason, the entropy and $\Delta$ are complementary for describing
the dynamical landscape (Figure 2C).
The two different modes in the $\Delta$-histogram are clearly reflected on
this 2D representation. Noticeably, the $\Delta$ values span a broad range
for any number of removed arrows, on the other hand higher entropies
are more frequent for larger number ($>2$) of removed arrows.
Qualitatively, the spread of points in the $H-\Delta$ plane conveys
a measure of \textit{robustness}. Accordingly, the $\Delta$ measure appears
more fragile than the entropy property, especially when few arrows are removed.

We now interpret the different locations in the  $H-\Delta$ plane:\\
1. If $\Delta$ is large ($\Delta\gtrsim 0.2$), the model does have attractor states which coincide with the gene activities
of the different cell-cycles phases. Such perturbations are specially 
interesting if the number of removed arrows is small (dark colors).
Such links are then essential for the model, as their removal disrupts the cell-cycle very efficiently.\\
2. If $\Delta$ is small and the entropy increases,
the probability that the dynamics ends in the reference attractor
decreases demonstrating that the removed arrows contributed
to the channeling properties of the system.\\
3. If $\Delta$ is small and the entropy decreases,
the main attractor of the perturbed network has a stronger attraction property.
Some of these networks could be considered as alternative cell-cycle models.\\

We illustrate these three regimes by examples:

\subsection{Examples of mutations}

In the first example (Table II), the dynamics has a large main basin of attraction like in the unperturbed model (Table I). However, 
the fixed point is significantly different from wild-type as
the system is blocked in the a state of the M-phase and
cannot finish properly the cell-cycle
(see Appendix for the recapitulation of the wild-type mode from \cite{tang_1}).

\begin{table}[h]
{\tiny
\begin{tabular}{cccccccccccl}
Cln3 & MBF & SBF & Cln1,2 & Cdh1 & Swi5 & Cdc20,14 & Clb5,6 & Sic1 & Clb1,2 & Mcm1 & $\%$ \\
0 & 0 & 0 & 0 & 1 & 0 & 0 & 0 & 1 & 0 & 0 & 0.8613 \\
0 & 0 & 1 & 1 & 0 & 0 & 0 & 0 & 0 & 0 & 0 & 0.0737 \\
0 & 1 & 0 & 0 & 1 & 0 & 0 & 0 & 1 & 0 & 0 & 0.0532 \\
0 & 0 & 0 & 0 & 0 & 0 & 0 & 0 & 1 & 0 & 0 & 0.0043 \\
0 & 0 & 0 & 0 & 0 & 0 & 0 & 0 & 0 & 0 & 0 & 0.0034 \\
0 & 1 & 0 & 0 & 0 & 0 & 0 & 0 & 1 & 0 & 0 & 0.0034 \\
0 & 0 & 0 & 0 & 1 & 0 & 0 & 0 & 0 & 0 & 0 & 0.0004 \\
\end{tabular}}
\caption{
Basins of attraction with their respective probabilities in ($\%$) for the original YCC network. The largest basin ends at the G1 stationary state. Entropy $H= 0.543$, Number of attractors $K=7$.}
\end{table}

\begin{table}
{\tiny
\begin{tabular}{cccccccccccl}
Cln3 & MBF & SBF & Cln1,2 & Cdh1 & Swi5 & Cdc20,14 & Clb5,6 & Sic1 & Clb1,2 & Mcm1 & $\%$ \\
0 & 0 & 0 & 0 & 0 & 1 & 1 & 0 & 1 & 1 & 1 & 0.880 \\
0 & 0 & 0 & 0 & 1 & 0 & 0 & 0 & 1 & 0 & 0 & 0.054 \\
0 & 0 & 1 & 1 & 0 & 0 & 0 & 0 & 0 & 0 & 0 & 0.027 \\
0 & 1 & 0 & 0 & 1 & 0 & 0 & 0 & 1 & 0 & 0 & 0.015 \\
0 & 0 & 0 & 0 & 1 & 1 & 1 & 0 & 1 & 1 & 1 & 0.010 \\
0 & 0 & 0 & 0 & 0 & 0 & 0 & 0 & 1 & 0 & 0 & 0.004 \\
0 & 0 & 0 & 0 & 0 & 0 & 0 & 0 & 0 & 0 & 0 & 0.003 \\
0 & 1 & 0 & 0 & 0 & 0 & 0 & 0 & 1 & 0 & 0 & 0.003 \\
0 & 0 & 0 & 0 & 1 & 0 & 0 & 0 & 0 & 0 & 0 & 0.000 \\
\end{tabular}}
\caption{
Basins of attraction with their respective probabilities, when (Cdc20,Cdc14) $\rightarrow$ Clb1,2 and Sic1 $\rightarrow$ Clb1,2 are removed. Entropy = 0.549, Number of attractors = 9, $\Delta$ = 0.41.}
\end{table}

In the second example (Table III), the dynamics has the same main fixed point as
the wild-type, but with a smaller basin of attraction, while the
second biggest has grown. Therefore the removed connection SBF
$\rightarrow$ Cln1,2 contributes to the ability of the main fixed point to funnel trajectories.

\begin{table}[h]
{\tiny
\begin{tabular}{cccccccccccl}
Cln3 & MBF & SBF & Cln1,2 & Cdh1 & Swi5 & Cdc20,14 & Clb5,6 & Sic1 & Clb1,2 & Mcm1 & $\%$\\
0 & 0 & 0 & 0 & 1 & 0 & 0 & 0 & 1 & 0 & 0 & 0.6669 \\
0 & 1 & 1 & 0 & 1 & 0 & 0 & 0 & 1 & 0 & 0 & 0.1762 \\
0 & 0 & 1 & 0 & 1 & 0 & 0 & 0 & 1 & 0 & 0 & 0.0654 \\
0 & 1 & 0 & 0 & 1 & 0 & 0 & 0 & 1 & 0 & 0 & 0.0532 \\
0 & 1 & 1 & 0 & 0 & 0 & 0 & 0 & 1 & 0 & 0 & 0.0180 \\
0 & 0 & 1 & 0 & 0 & 0 & 0 & 0 & 1 & 0 & 0 & 0.0043 \\
0 & 0 & 0 & 0 & 0 & 0 & 0 & 0 & 1 & 0 & 0 & 0.0043 \\
0 & 0 & 0 & 0 & 0 & 0 & 0 & 0 & 0 & 0 & 0 & 0.0034 \\
0 & 0 & 1 & 0 & 0 & 0 & 0 & 0 & 0 & 0 & 0 & 0.0034 \\
0 & 1 & 0 & 0 & 0 & 0 & 0 & 0 & 1 & 0 & 0 & 0.0034 \\
0 & 0 & 0 & 0 & 1 & 0 & 0 & 0 & 0 & 0 & 0 & 0.0004 \\
0 & 0 & 1 & 0 & 1 & 0 & 0 & 0 & 0 & 0 & 0 & 0.0004 \\
\end{tabular}}
\caption{
Basins of attraction with their respective probabilities, when SBF $\rightarrow$ Cln1,2 is removed. Entropy $H=1.096$, Number of attractors $K=12$, $\Delta$ = 0.05.}
\end{table}

The third example (Table IV) is a model with four removed arrows which has the same
main fixed point with a slightly higher probability. Also, the second 
largest fixed point is same as in the wild-type model. This indicates that the effect
of some mutations can be canceled by further mutations.
While such cases exist, we found that networks with several removed links that
preserving the unperturbed cell-cycle behavior are rare.

\begin{table}[h]
{\tiny \vspace{0.5cm}
\begin{tabular}{cccccccccccl}
Cln3 & MBF & SBF & Cln1,2 & Cdh1 & Swi5 & Cdc20,14 & Clb5,6 & Sic1 & Clb1,2 & Mcm1 & $\%$\\
0 & 0 & 0 & 0 & 1 & 0 & 0 & 0 & 1 & 0 & 0 & 0.8793 \\
0 & 0 & 1 & 1 & 0 & 0 & 0 & 0 & 0 & 0 & 0 & 0.0507 \\
0 & 1 & 0 & 0 & 1 & 0 & 0 & 0 & 1 & 0 & 0 & 0.0356 \\
0 & 0 & 0 & 0 & 0 & 0 & 0 & 0 & 1 & 0 & 0 & 0.0268 \\
0 & 1 & 0 & 0 & 0 & 0 & 0 & 0 & 1 & 0 & 0 & 0.0034 \\
0 & 0 & 0 & 0 & 0 & 0 & 0 & 0 & 0 & 0 & 0 & 0.0034 \\
0 & 0 & 0 & 0 & 1 & 0 & 0 & 0 & 0 & 0 & 0 & 0.0004 \\
\end{tabular}}
\caption{
Basins of attraction with their respective probabilities, when (Cdc20,Cdc14) $\rightarrow$ Clb1,2, Clb1,2 $\rightarrow$ Mcm1, Clb1,2 $\rightarrow$ Cdh1 and Clb1,2 $\rightarrow$ Swi5 are removed.
Entropy $H= 0.523$, Number of attractors $K=7$, $\Delta$ = 0.025.}
\end{table}

\vspace{-0.5cm}
\section{Conclusion}

We have proposed a systematic approach for studying the dynamical
attractor landscape of biological networks, and their response to structural
perturbations. In particular, we introduced a low dimensional representation
of the system of attractors, the entropy, and a probabilistic measure
in the perturbation size $\Delta$.
This enabled us to study the global characteristics of network
perturbation in a compact and visually effective form.
In a biological context, this can provide hints to elucidate the dynamical role of
specific network links. Alternatively, the function of new and yet unobserved links can
be predicted as in \cite{stoll}, and imperfect starting models can be improved.

We applied this method to a model of the yeast cell-cycle by Li et al.
Using the measures introduced, we have generalized the dynamical characterization of
the model using a broad range of perturbations.
This has enabled us to emphasize the breadth of
dynamical behavior (Figure 2) induced by only few mutated links.
Interestingly, 
we observed (Figure 2C) that the structure of the system of attractors ($H$) behaves quite robustly
compared to the modification in the final states ($\Delta$), especially when the number of removed links
is small ($<3$).
We illustrated through examples the consequences of removing
individual or groups of links. Interestingly it was possible to remove up to four   links while
not affecting the basin structure significantly. 
Tracking the dynamical changes in the activity levels of proteins in a network is a very high-dimensional problem. 
It therefore important to be have few informative variables which allow one to efficiently assess
a large number of perturbed models at once.
We believe that basin entropy and distance to a reference attractor are well suited for this purpose.

\subsection*{Acknowledgments}
We thank the organizers of the CompBioNets '04 conference (Recife, Brazil) at which an initial version of this work was presented. The simulations were performed on an Itanium2 cluster from HP/Intel at the Vital-IT facilities.
FN ad GS acknowledge funding from the NCCR Molecular Oncology program and NIH administrative supplement to parent
grant GM54339.

\appendix
\section{the Yeast cell-cycle network of Li et al.}
The following two tables are recapitulated from \cite{tang_1}.

\begin{table}[h]
{\tiny
\begin{tabular}{cccccccccccc}
1$\stackrel{+}{\rightarrow}$2&
1$\stackrel{+}{\rightarrow}$3&
2$\stackrel{+}{\rightarrow}$8&
3$\stackrel{+}{\rightarrow}$4&
6$\stackrel{+}{\rightarrow}$9&
7$\stackrel{+}{\rightarrow}$5&
7$\stackrel{+}{\rightarrow}$6&
7$\stackrel{+}{\rightarrow}$9&
8$\stackrel{+}{\rightarrow}$10&
8$\stackrel{+}{\rightarrow}$11&
10$\stackrel{+}{\rightarrow}$7&
10$\stackrel{+}{\rightarrow}$11\\

11$\stackrel{+}{\rightarrow}$6&
11$\stackrel{+}{\rightarrow}$7&
11$\stackrel{+}{\rightarrow}$10&
4$\stackrel{-}{\rightarrow}$9&
4$\stackrel{-}{\rightarrow}$5&
5$\stackrel{-}{\rightarrow}$10&
7$\stackrel{-}{\rightarrow}$8&
7$\stackrel{-}{\rightarrow}$10&
8$\stackrel{-}{\rightarrow}$5&
8$\stackrel{-}{\rightarrow}$9&
9$\stackrel{-}{\rightarrow}$8&
9$\stackrel{-}{\rightarrow}$10\\

10$\stackrel{-}{\rightarrow}$2&
10$\stackrel{-}{\rightarrow}$3&
10$\stackrel{-}{\rightarrow}$5&
10$\stackrel{-}{\rightarrow}$6&
10$\stackrel{-}{\rightarrow}$9&
1$\stackrel{-}{\rightarrow}$1&
4$\stackrel{-}{\rightarrow}$4&
6$\stackrel{-}{\rightarrow}$6&
7$\stackrel{-}{\rightarrow}$7&
11$\stackrel{-}{\rightarrow}$11\\
\end{tabular}}
\caption{Adjacency matrix of the Yeast cell-cycle network. The numbers refer to the ordering of the nodes as used in Tables I-IV,VI. $+$ (respectively $-$)
represent activating (respectively repressing) links.  }
\end{table}


\begin{table}[h]
{\tiny
\begin{tabular}{ccccccccccccc}
 t & Cln3 & MBF & SBF & Cln1,2 & Cdh1 & Swi5 & C20,14 & Clb5,6 & Sic1 & Clb1,2 & Mcm1 & Phase \\ \\
1  & 1  & 0  & 0  & 0  & 1  & 0  & 0  & 0  & 1  & 0  & 0  & START \\
2  & 0  & 1  & 1  & 0  & 1  & 0  & 0  & 0  & 1  & 0  & 0  & G1 \\
3  & 0  & 1  & 1  & 1  & 1  & 0  & 0  & 0  & 1  & 0  & 0  & G1 \\
4  & 0  & 1  & 1  & 1  & 0  & 0  & 0  & 0  & 0  & 0  & 0  & G1 \\
5  & 0  & 1  & 1  & 1  & 0  & 0  & 0  & 1  & 0  & 0  & 0  & S \\
6  & 0  & 1  & 1  & 1  & 0  & 0  & 0  & 1  & 0  & 1  & 1  & G2 \\
7  & 0  & 0  & 0  & 1  & 0  & 0  & 1  & 1  & 0  & 1  & 1  & M \\
8  & 0  & 0  & 0  & 0  & 0  & 1  & 1  & 0  & 0  & 1  & 1  & M \\
9  & 0  & 0  & 0  & 0  & 0  & 1  & 1  & 0  & 1  & 1  & 1  & M \\
10  & 0  & 0  & 0  & 0  & 0  & 1  & 1  & 0  & 1  & 0  & 1  & M \\
11  & 0  & 0  & 0  & 0  & 1  & 1  & 1  & 0  & 1  & 0  & 0  & M \\
12  & 0  & 0  & 0  & 0  & 1  & 1  & 0  & 0  & 1  & 0  & 0  & G1 \\
13 & 0 & 0 & 0 & 0 & 1 & 0 & 0 & 0 & 1 & 0 & 0 & G1* \\
\end{tabular}}
\caption{This table represents the discrete time evolution of the boolean states of the YCC network as it traverses the different cell-cycle phases. Cdc20.14 has been abbreviated C20,14; G1* indicates the stationary G1 phase. }
\end{table}

\end{document}